\documentclass{jfm}
\usepackage{amsmath}
\usepackage{upgreek}
\usepackage{graphicx}
\usepackage{epstopdf, epsfig}
\usepackage{subfigure}

\newcommand{\dee}{\mathrm{d}}
\newcommand{\pee}{\mathbf{P}}
\newcommand{\tee}{\mathbf{T}}
\newcommand{\ohm}{\boldsymbol{\omega}}
\newcommand{\Ohm}{\boldsymbol{\Omega}}
\newcommand{\you}{\mathbf{U}}
\newcommand{\yoo}{\mathbf{u}}
\newcommand{\ecks}{\mathbf{x}}
\newcommand{\en}{\mathbf{n}}
\newcommand{\eee}{\mathbf{e}}
\newcommand{\eff}{\mathbf{F}}
\newcommand{\gam}{\boldsymbol{\gamma}}
\newcommand{\dx}{\,\dee\ecks}
\newcommand{\dtwox}{\,\dee^2\ecks}
\newcommand{\dthreex}{\,\dee^3\ecks}

\shorttitle{Force decomposition for incompressible flow}
\shortauthor{W.~R.~Graham}

\title{Decomposition of the forces on a body moving in an incompressible fluid}

\author{W.~R.~Graham\aff{1}
  \corresp{\email{wrg11@cam.ac.uk}}}

\affiliation{\aff{1}Department of Engineering, University of
Cambridge, Trumpington Street, Cambridge CB2 1PZ, UK}

\begin{document}

\maketitle

\begin{abstract}
In analysing fluid forces on a moving body, a natural approach is to seek a component due to viscosity and an `inviscid' remainder.  It is also attractive to decompose the velocity field into irrotational and rotational parts, and apportion the force resultants accordingly.  The `irrotational' resultants can then be identified as classical `added mass', but the remaining, `rotational', resultants appear not to be consistent with the physical interpretation of the rotational velocity field (as that arising from the fluid vorticity with the body stationary).  The alternative presented here splits the inviscid resultants into components that are unquestionably due to independent aspects of the problem: `convective' and `accelerative'.  The former are associated with the pressure field that would arise in an inviscid flow with (instantaneously) the same velocities as the real one, and with the body's velocity parameters --- angular and translational --- unchanging.  The latter correspond to the pressure generated when the body accelerates from rest in quiescent fluid with its given rates of change of angular and translational velocity.  They are reminiscent of the added-mass force resultants, but are simpler, and closer to the standard rigid-body inertia formulae, than the developed expressions for added-mass force and moment.  Finally, the force resultants due to viscosity also include a contribution from pressure.  Its presence is necessary in order to satisfy the equations governing the pressure field, and it has previously been recognised in the context of `excess' stagnation-point pressure.  However, its existence does not yet seem to be widely appreciated.
\end{abstract}

\section{Introduction} \label{secintro}
One of the most fundamental problems in fluid mechanics is the determination of the force resultants on an immersed body.  When the body is stationary, it experiences a hydrostatic force which is easily calculated.  When it moves, however, the situation is much more complex.  Although the governing equations have long been well established, solving them remains, in general, extremely difficult.  Thus we still seek simplified conceptual representations, both for estimation of force resultants and for physical insight.  In particular, it is desirable to decompose the force resultants into components that can be associated with separate aspects of the fluid flow.

The velocity field in the fluid admits a straightforward decomposition into irrotational (`potential') and rotational (`circulatory') components \citep[cf., for example, ][\S2.4]{batchelorFluidDyn}.  Hence it is natural to seek a corresponding decomposition for the force resultants.  As noted by \citet{changforces}, this approach also has the advantage that an extensive established theory for the potential-flow forces can be exploited.  It has subsequently been adopted by \citet{howeOnForce}, \citet{eldredgeReconciliation}, and \citet{limachergenderiv}.

The other obvious partition arises from the form of the stress field in the fluid, which has contributions from pressure and from viscous stresses.  Thus one can identify separate force resultants associated with the viscous and inviscid aspects of the flow.  Indeed, given that the potential theory is predicated on inviscid fluid behaviour, this decomposition can be applied in conjunction with a split into potential and circulatory components.  Of the formulations cited above, those of \citet{changforces}, \citet{howeOnForce}, and \citet{eldredgeReconciliation} explicitly isolate viscous contributions.

There is, however, an undesirable feature of the potential/circulatory force-resultant decomposition: the formulae describing the circulatory part include contributions from the potential flow.  This is only made explicit by \citet{changforces}, but seems, in fact, to be general.  The reason will be explained later; the implication is that the decomposition may not be optimal.  Specifically, one can ask whether an alternative, associated with independent physical processes, can be found.  This is the topic of the current work.

\begin{figure}
\begin{center}
\includegraphics{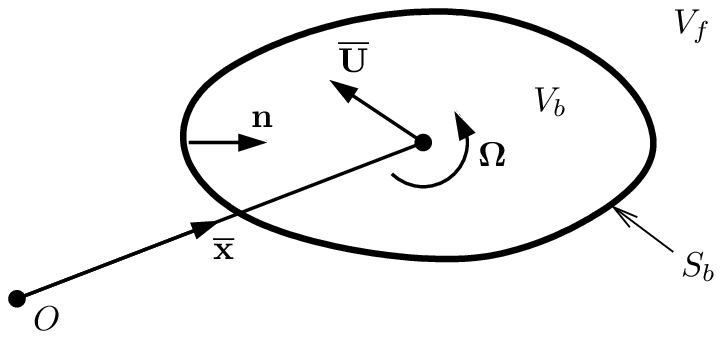}
\caption{{\label {config}} A body moving with velocity $\overline{\you}$ and angular velocity $\Ohm$ in an incompressible fluid.}
\end{center}
\end{figure}

The nomenclature defining the (rigid) body and its motion is introduced in Fig.~\ref{config}.  In a fixed frame of reference with origin $O$, the body's centre of volume is at position $\ecks = \overline{\ecks}$, moving with velocity $\overline{\you}$.  The body is also rotating, with angular velocity $\Ohm$.  The region inside the body is denoted $V_{b}$, and that outside $V_{f}$; their common boundary is the body surface, $S_{b}$.  The unit vector normal to this surface, pointing into the body, is denoted $\en$.  The fluid occupying $V_{f}$ has constant density $\rho$, dynamic viscosity $\mu$, and kinematic viscosity $\nu = \mu/\rho$, and is at rest at infinity.  Its pressure (less the hydrostatic component) and velocity will be represented by $p$ and $\yoo$ respectively.

For the body's velocity away from its centre of volume, we use the standard result that, if $\ecks$ is the current position of a point in the body, then the rate of change of the body-fixed vector that instantaneously coincides with $\ecks-\overline{\ecks}$ is $\Ohm\times(\ecks-\overline{\ecks})$.  Hence the velocity $\you$ of the body point is given by

\begin{equation}
\you = \overline{\you} + \Ohm\times\left(\ecks-\overline{\ecks}\right) .
\end{equation}
Similarly, its acceleration is

\begin{equation}
\dot{\you} = \frac{\dee\overline{\you}}{\dee t} + \frac{\dee\Ohm}{\dee t}\times\left(\ecks-\overline{\ecks}\right) + \Ohm \times \left[\Ohm\times\left(\ecks-\overline{\ecks}\right) \right].
\label{eqUdot}
\end{equation}
The dot notation employed here will be reserved for body-fixed time derivatives.

Also needed subsequently will be the curl of $\you$, which can straightforwardly be evaluated as $2\Ohm$.  Unlike the body and fluid velocities, which (by virtue of the no-slip condition) are continuous at the body surface, the respective curls need not match.  Only continuity of the normal components is guaranteed by the no-slip condition.

The force-resultant decomposition presented in this paper arises from a partition of the pressure field into independent `viscous', `convective', and `accelerative' components.  This is presented in the following section.  Next, in \S\ref{seccompvm}, we demonstrate that it is consistent with the viscous/inviscid decomposition derived \citep[from the vorticity-moment representation of][]{wuTheory} by \citet{eldredgeReconciliation}.  Readers prepared to accept the legitimacy of the current formulation without further proof can skip ahead to the final part of the paper, \S\ref{secimp4am}, in which the force resultants arising from the accelerative pressure component are compared with those in the classical added-mass formulae for potential flow.

\section{The pressure-field decomposition}
The most direct formulation for the forces on the body is in terms of the surface traction, which is made up of a pressure and a viscous-stress contribution.  The latter is uniquely associated with the fluid's viscosity, but the pressure arises from several influences.  They can be identified from the governing equations for the pressure field, which are derived in \S\ref{secgoveq}.  Arising naturally from these equations is a viscous/inviscid decomposition (\S\ref{secviscinv}); further breakdown of the inviscid part into convective and accelerative components is described subsequently in \S\ref{secinvcomp}.

\subsection{Governing equations}\label{secgoveq}
The fluid flow is described by the continuity equation,

\begin{equation}
\nabla\cdot\yoo=0,
\label{eqcont}
\end{equation}
\noindent and the Navier-Stokes equation of motion,

\begin{equation}
\rho\left(\frac{\partial\yoo}{\partial t}+\yoo\cdot\nabla\yoo\right)=-\nabla p + \mu\nabla^{2}\yoo .
\end{equation}
Here the viscous contribution can alternatively be written in terms of the vorticity, $\ohm = \nabla\times\yoo$, via the identity $\nabla^{2}\yoo = \nabla(\nabla\cdot\yoo) -\nabla\times\ohm$.  In the light of (\ref{eqcont}), we have

\begin{equation}
\nabla^{2}\yoo = -\nabla\times\ohm ,
\label{eqd2ueqcu}
\end{equation}
and hence

\begin{equation}
\rho\left(\frac{\partial\yoo}{\partial t}+\yoo\cdot\nabla\yoo\right)=-\nabla p - \mu\nabla\times\ohm .
\label{eqns}
\end{equation}

Given the velocity as a function of position and time, this is sufficient to define the pressure field.  However, a more useful formulation follows from taking the divergence of (\ref{eqns}), which yields

\begin{equation}
\nabla^{2}p=-\rho\nabla\cdot(\yoo\cdot\nabla\yoo).
\label{eqd2p}
\end{equation}

\noindent To specify the pressure uniquely, this Poisson equation also requires Dirichlet or Neumann boundary conditions at infinity (where $p \rightarrow 0$) and at the body surface. Here the Neumann form is appropriate, since it follows directly from the Navier-Stokes equation; we have
\begin{equation}
\en\cdot\nabla p =-\mu\en\cdot\nabla\times\ohm-\rho\en\cdot\dot{\you}.
\label{eqpbc}
\end{equation}

\noindent (Note that the no-slip condition has been invoked to replace the material derivative of the fluid velocity, $\partial\yoo/\partial t+\yoo\cdot\nabla\yoo$, with the body acceleration $\dot{\you}$.)

A notable feature of (\ref{eqd2p}) and (\ref{eqpbc}) is that time dependence only enters explicitly via the body acceleration.  Hence, given an instantaneous fluid-velocity field and known body motion, the pressure is specified.  Equivalently, we can say that, unlike the velocity field, the pressure field carries complete knowledge of the body's acceleration.

Also relevant is the term in (\ref{eqpbc}) proportional to viscosity.  This shows that the viscous contribution to the force on the body does not arise from viscous stresses alone; viscosity affects the pressure field too.  Indeed, this field can be explicitly decomposed into unique `inviscid' and `viscous' components, as will now be shown.  

\subsection{Viscous/inviscid decomposition}\label{secviscinv}
Consider an inviscid fluid, subject to the same body motion, and with, instantaneously, the same velocity field as the true, viscous, flow.  The associated pressure field would satisfy the viscosity-independent governing equation, (\ref{eqd2p}), but not the boundary condition, (\ref{eqpbc}).  Hence there must be, in the true flow, an additional `viscous' pressure component.

With this point in mind, we observe that (\ref{eqd2p}) and (\ref{eqpbc}) admit the decomposition $p=p^{(i)}+p^{(v)}$, in which $p^{(i)}$ --- the inviscid component --- satisfies
\begin{equation}
\nabla^{2}p^{(i)}=-\rho\nabla\cdot\left(\yoo\cdot\nabla\yoo\right)
\label{eqd2pi}
\end{equation}

\noindent in the fluid and
\begin{equation}
\en\cdot\nabla p^{(i)}=-\rho\en\cdot\dot{\you}
\label{eqpibc}
\end{equation}

\noindent on the body, while $p^{(v)}$ --- the viscous component --- has 

\begin{equation}
\nabla^{2}p^{(v)} = 0
\label{eqd2pv}
\end{equation}
\noindent in the fluid and 

\begin{equation}
\en\cdot\nabla p^{(v)} = -\mu\en\cdot\nabla\times\ohm
\label{eqpvbc}
\end{equation}

\noindent on the body.  This split makes $p^{(v)}$ directly proportional to the fluid viscosity, so may appear self-evident.  However, some further analysis is needed to show that it does indeed correspond to the foregoing qualitative description.

The problem lies in the boundary condition for $p^{(i)}$, which implicitly incorporates the no-slip condition.  The pressure field associated with the putative instantaneous inviscid flow has normal gradient $-\rho\en\cdot\left(\partial\yoo/\partial t + \yoo\cdot\nabla\yoo\right)$ at the boundary, and this might differ from $-\rho\en\cdot\dot{\you}$, even though the flow and body velocities match at this instant.  In fact, however, it does not, as can be shown by differentiating the inviscid-flow no-penetration condition, $\yoo\cdot\en=\you\cdot\en$, with respect to time at a fixed point on the body.  This yields

\begin{equation}
\en\cdot\left(\frac{\partial\yoo}{\partial t}+\yoo\cdot\nabla\yoo\right)=\en\cdot\dot{\you}+\en\cdot\left[(\yoo-\you) \cdot\nabla\yoo\right]-(\Ohm\times\en)\cdot(\yoo-\you)
\end{equation}
in general.  In our case, though, with $\yoo = \you$ instantaneously, the second and third terms on the right-hand side disappear.  Hence $p^{(i)}$ as specified by (\ref{eqd2pi}) and (\ref{eqpibc}) is indeed the pressure field that would be found in the matching-velocity inviscid flow, and $p^{(v)}$ can unambiguously be identified as the additional component due to viscosity.

Note, finally, that there must be no confusion between the `viscous' pressure and the normal component of the viscous-stress tensor.  The component $p^{(v)}$ represents a pressure arising from the need to balance viscous stresses, not a viscous stress itself.  Its contribution is evident in the analyses of total pressure in viscous flow conducted by \citet{issaTotp} and \citet{williamsTotp}, both of whom show that the pressure at a stagnation point can exceed the total pressure `at infinity' by an amount proportional to the viscosity.  In Appendix \ref{appvpegs}, we show that $p^{(v)}$ can be explicitly identified in the pressure fields of exact analytical solutions to the Navier-Stokes equation.  In particular, in the classical parallel flow of Poiseuille, it is responsible for the constant streamwise pressure gradient, with $p^{(i)}$ zero throughout.  The existence of $p^{(v)}$ is also implicit in the formula for pressure force given by \citet{changforces}, which includes a term proportional to viscosity.  Nonetheless, its contribution does not appear in the force breakdown proposed by \cite{eldredgeReconciliation}.  This issue will be addressed in \S\ref{seccompvm}.  For the moment, we continue to further decomposition of $p^{(i)}$.

\subsection{Decomposition of the inviscid-pressure field}\label{secinvcomp}

\subsubsection{The potential/circulatory decomposition}
First, we are now in a position to substantiate the claim made in \S\ref{secintro}: that the circulatory part of the potential/circulatory force-resultant decomposition has a dependence on the potential part of the flow field.  The potential/circulatory velocity-field decomposition is $\yoo = \yoo_{\phi} + \yoo_{\omega}$, where $\yoo_{\phi}$ represents irrotational flow around the moving body, and $\yoo_{\omega}$ flow of vorticity $\nabla\times\yoo$ with the body stationary at its current position.  Under this decomposition, the right-hand side of (\ref{eqd2pi}) becomes $-\rho\nabla\cdot\left[ (\yoo_{\phi} + \yoo_{\omega})\cdot\nabla(\yoo_{\phi} + \yoo_{\omega}) \right]$.  Of the terms that arise on its expansion, $-\rho\nabla\cdot (\yoo_{\phi}\cdot\nabla\yoo_{\phi})$ corresponds to the potential-flow pressure field, and the remainder (by definition) to the circulatory component.  Included in the remainder are products involving $\yoo_{\phi}$ and $\yoo_{\omega}$, so, unless their contribution is somehow identically zero, the irrotational velocity field influences the circulatory force resultants.  It then follows that these resultants differ from those that would arise in the physical situation corresponding to the rotational velocity component, i.e.~flow $\yoo_{\omega}$ with the body fixed.  Equally, the potential force resultants do not fully encapsulate the effects of the body's motion.

In the light of this argument, it should be possible to identify an implicit co-dependence in the nominally independent potential and circulatory force resultants of \citet{howeOnForce}, \citet{eldredgeReconciliation}, and \citet{limachergenderiv}.  This is most straightforward for the Howe formulation.  Here the circulatory force resultants contain integrals involving the cross-product of the vorticity and velocity fields.  Crucially, the latter includes the potential contribution, so the irrotational velocity component affects the circulatory force resultants.

The remaining formulations represent the circulatory force resultants in terms of rates of change of vorticity-moment integrals.  As these integrals depend on the vorticity field alone, the influence of the potential velocities is not immediately evident.  However, it appears when the time derivative is applied.  Thus, for example, the circulatory element of the force decomposition contains the term

\begin{equation}
\frac{\dee}{\dee t} \int_{V_f} \ecks\times\ohm \,\dee V = \int_{V_f}\yoo\times\ohm \,\dee V + \int_{V_f}\ecks\times\left(\frac{\partial\ohm}{\partial t}+\yoo\cdot\nabla\ohm\right) \,\dee V ,
\end{equation}
where the right-hand side follows by applying the derivative to the sum of infinitesimal material elements implied by the volume integral.  The integral on the left-hand side is independent of the potential velocity field, but its rate of change is not.  Again, this implies that the circulatory force is not independent of the body's motion.

\subsubsection{The convective/accelerative decomposition}
An alternative decomposition which \emph{does} lead to physically independent components arises from the observation that only the boundary condition, (\ref{eqpibc}), contains information on the rates of change of the body's translational and angular velocities, $\dee\overline{\you}/\dee t$ and $\dee\Ohm/\dee t$.  It is thus possible to split the inviscid-pressure field into a component associated with the instantaneous fluid and body velocities, and a component associated with $\dee\overline{\you}/\dee t$ and $\dee\Ohm/\dee t$ alone.  These quantities will be referred to as the `convective' and `accelerative' pressure fields ($p^{(c)}$ and $p^{(a)}$).

Under the decomposition $p^{(i)} = p^{(c)} + p^{(a)}$, the Poisson equation for $p^{(i)}$, (\ref{eqd2pi}), yields

\begin{equation}
\nabla^{2}p^{(c)}=-\rho\nabla\cdot\left(\yoo\cdot\nabla\yoo\right)
\label{eqpc}
\end{equation}
and

\begin{equation}
\nabla^{2}p^{(a)} = 0.
\label{eqpa}
\end{equation}
The corresponding boundary conditions are, via (\ref{eqpibc}) and (\ref{eqUdot}),

\begin{equation}
\en\cdot\nabla p^{(c)} = -\rho\en\cdot\left\{ \Ohm \times \left[ \Ohm\times\left(\ecks-\overline{\ecks}\right) \right] \right\}
\end{equation}
and

\begin{equation}
\en\cdot\nabla p^{(a)} = -\rho\en\cdot\left\{ \frac{\dee\overline{\you}}{\dee t} + \frac{\dee\Ohm}{\dee t}\times\left(\ecks-\overline{\ecks}\right) \right\}.
\label{eqpabc}
\end{equation}
The interpretation of the two inviscid-pressure components is straightforward.  The `accelerative' element is the field that would arise if the body and fluid were both initially stationary, and the body were then instantaneously given translational acceleration $\dee\overline{\you}/\dee t$ and angular acceleration $\dee\Ohm/\dee t$.  The `convective' element is the field that would be observed in inviscid flow with the same instantaneous body and fluid velocities as the real case, and with the body's translational and angular velocities unchanging.  Remarkably, the pressure for the general situation, in which the two motions are combined, is simply the sum of the pressures for the individual cases.

Associated with $p^{(c)}$ and $p^{(a)}$ are similarly independent components of the force on the body.  It is tempting to call the second `added mass', since it is linked to (part of) the body's acceleration.  This, however, would be inconsistent with the classical definition, in which the term refers to the force resultants in the full potential flow implied by the body motion.  The difference, and its implications, will be explored in \S\ref{secimp4am}.

\section{Comparison with the vorticity-moment formulation}\label{seccompvm}
The viscous/inviscid decomposition presented in \S\ref{secviscinv} implies that the viscous contribution to the forces on the body arises not only from the viscous stress, but also the pressure component $p^{(v)}$.  This conclusion contradicts \citet{eldredgeReconciliation}, who derives a pure viscous-stress contribution from the vorticity-moment expression for the body force \citep{wuTheory}.  To resolve the issue, it is necessary to reconcile the vorticity-moment (or `impulse') formulation with the surface-traction forces arising from $p^{(i)}$, $p^{(v)}$, and the viscous stress.  This is the topic of the current section.

The decomposition based on vorticity moment involves vortex sheets, despite the no-slip condition, because the inviscid and viscous components separately admit the development of a tangential velocity discontinuity at the body surface.  First, then, we derive expressions for the rates of growth of these sheets.  Next, in \S\ref{secstvm2d}, we address the reconciliation issue in the context of the resultant force on the body in the two-dimensional case.  This differs sufficiently from its three-dimensional counterpart to warrant separate treatment, and is considered first because it is the configuration discussed by \citet{eldredgeReconciliation}.  The corresponding analysis in three dimensions follows in \S\ref{secstvm3d}.  Similar procedures confirm consistency between the surface-traction and vorticity-moment expressions for the resultant moment on the body.  These are not documented here, because they provide no further conceptual illumination.

The manipulations required in this section involve extensive use of Gauss's theorem and its variants \citep[cf., for example,][\S1.4.2]{zangwillElectroDyn}.  These can conveniently be summarised, via index notation, in the form

\begin{equation}
\int_{V} \frac{\partial f}{\partial x_{i}} \,\dee V = \int_{S} f n_{i} \,\dee S ,
\end{equation}
where $f(\ecks)$ can be a scalar or a component of a higher-dimensional entity.  The volume $V$ should, in principle, be bounded by the surface(s) $S$, but infinite domains are permissible if the surface integral `at infinity' tends to zero.  Subject to the relevant conditions on velocity and vorticity there \citep{wuTheory}, this is true for the instances arising below.  Finally, to distinguish the two- and three-dimensional configurations, the volume (surface) element will be represented by $\dtwox$ ($\dx$) for the former, and $\dthreex$ ($\dtwox$) for the latter.

\subsection{Vortex sheets}
Consider a boundary where the no-penetration condition, $\yoo\cdot\en=\you\cdot\en$, applies, but where there is slip.  The slip corresponds to a vortex sheet whose strength is given by

\begin{equation}
\gam=-\en\times(\yoo-\you) .
\end{equation}
Differentiating this expression with respect to time at a fixed point on the body surface yields

\begin{equation}
\dot{\gam}=-\en\times\left(\frac{\partial\yoo}{\partial t}+\you\cdot\nabla\yoo-\dot{\you}\right)-(\Ohm\times\en)\times(\yoo-\you) .
\label{eqgamdot}
\end{equation}
In our full, viscous, flow, with $\yoo = \you$ and $\partial\yoo/\partial t+\yoo\cdot\nabla\yoo = \dot{\you}$, both $\gam$ and $\dot{\gam}$ are identically zero, but the same cannot necessarily be said of the separate components we have identified.  Indeed, the viscous/inviscid decomposition of \S\ref{secviscinv} corresponds to a split of $\dot{\gam}$ into two mutually cancelling non-zero components, $\dot{\gam}^{(i)}$ and $\dot{\gam}^{(v)}$, as follows.

The inviscid flow associated with $p^{(i)}$ has $\gam = 0$ (by definition, because its velocity instantaneously satisfies the no-slip condition).  However, $p^{(i)}$ also contains information about the velocity field's rate of change.  Specifically, the governing (Euler) equation implies an inviscid component of $\partial\yoo/\partial t$ given by

\begin{equation}
\left(\frac{\partial\yoo}{\partial t}\right)^{(i)} = -\yoo\cdot\nabla\yoo - \frac{1}{\rho} \nabla p^{(i)} .
\end{equation}
The associated vortex-sheet growth rate is found by replacing $\partial\yoo/\partial t$ in (\ref{eqgamdot}) with $(\partial\yoo/\partial t)^{(i)}$, and recalling that the matching inviscid flow has $\yoo = \you$ on the body surface at this instant.  Hence

\begin{equation}
\dot{\gam}^{(i)} = \en\times\left(\frac{1}{\rho} \nabla p^{(i)} + \dot{\you}\right) .
\label{eqgdi}
\end{equation}
Now the viscous vortex-sheet growth rate follows by writing $\dot{\gam}^{(v)} = \dot{\gam} - \dot{\gam}^{(i)}$ and invoking (\ref{eqgamdot}), (\ref{eqns}) and (\ref{eqgdi}).  The result is

\begin{equation}
\dot{\gam}^{(v)}=\en\times\left(\frac{1}{\rho}\nabla p^{(v)}+\nu\nabla\times\ohm\right) .
\label{eqgdv}
\end{equation}

\subsection{Surface-traction/vorticity-moment comparison (two-dimensional)} \label{secstvm2d}
We begin with the vorticity-moment formulation \citep{wuTheory}.  Here, the force exerted by the fluid on the body is given by

\begin{equation}
\eff=\rho V_{b}\frac{\dee\overline{\you}}{\dee t} - \rho \frac{\dee\boldsymbol{\alpha}}{\dee t},
\label{eqFWu2d}
\end{equation}
where $\boldsymbol{\alpha}$ is the first moment of the combined fluid/body vorticity field:

\begin{equation}
\boldsymbol{\alpha} = \int_{V_b}\ecks\times(2\Ohm) \dtwox + \int_{V_f}\ecks\times\ohm \dtwox.
\label{eqalfa2d}
\end{equation}
As an aside, note that the force is not evidently origin-independent (as it must be) in this formulation.  It is possible to manipulate (\ref{eqalfa2d}) and the results we derive from it to make this property explicit.  However, for our purposes, there is no obvious benefit in doing so.

The first integral in (\ref{eqalfa2d}) is $-2V_{b}\Ohm\times\overline{\ecks}$, and the time differentiation required in (\ref{eqFWu2d}) can be carried out straightforwardly.  For the second, the differential must be taken inside the volume integral, and this can be done in various ways.  Here, we consider the integral as over material fluid elements, with constant $\dee^{2}\ecks$ because of incompressibility.  The time variation of each term in the integrand is then given by application of the convective derivative, $\partial/\partial t+\yoo\cdot\nabla$.  Thus the time derivative of (\ref{eqalfa2d}) is

\begin{equation}
\frac{\dee\boldsymbol{\alpha}}{\dee t} = -2V_{b}\left(\Ohm\times\overline{\you} + \frac{\dee\Ohm}{\dee t}\times\overline{\ecks} \right) + \int_{V_f}\yoo\times\ohm \dtwox + \int_{V_f}\ecks\times\left(\frac{\partial\ohm}{\partial t}+\yoo\cdot\nabla\ohm\right)\dtwox .
\label{eqdalphadt}
\end{equation}
The second term can be manipulated into an integral on the body surface via the standard identity $\yoo\times\ohm = \nabla(\frac{1}{2}\yoo\cdot\yoo) - \yoo\cdot\nabla\yoo$, the continuity condition $\nabla\cdot\yoo = 0$, and invocation of  Gauss's theorem.  The integrand can then be expressed in terms of the body kinematics, whereupon a further application of Gauss's theorem leads to a body-volume integral, which can be evaluated  to yield

\begin{equation}
\int_{V_f}\yoo\times\ohm \dtwox = 2V_{b}\Ohm\times\overline{\you} .
\label{eqintVf}
\end{equation}
(This is consistent with a result given by \citet[\S3.2]{saffman}, namely that $\yoo\times\ohm$ for the combined body/fluid velocity field has zero volume integral.)  Hence the vorticity-moment expression for the force on the body, (\ref{eqFWu2d}), becomes

\begin{equation}
\eff=\rho V_{b}\frac{\dee\overline{\you}}{\dee t} + 2\rho V_{b}\frac{\dee\Ohm}{\dee t}\times\overline{\ecks} - \rho \int_{V_f}\ecks\times\left(\frac{\partial\ohm}{\partial t}+\yoo\cdot\nabla\ohm\right)\dtwox .
\label{eqFtwod}
\end{equation}

As it stands, this expression is unsuitable for decomposition into inviscid and viscous components, because it does not include the vortex-sheet growth present in each.  However, since the overall growth rate $\dot{\gam}$ is zero, its first moment can be added without altering $\eff$.  We also use the two-dimensional vorticity equation,

\begin{equation}
\frac{\partial\ohm}{\partial t}+\yoo\cdot\nabla\ohm = \nu \nabla^{2} \ohm ,
\end{equation}
in the final term of (\ref{eqFtwod}).  The upshot is the force decomposition

\begin{equation}
\eff=\eff^{(i)}+\eff^{(v)},
\end{equation}
with

\begin{equation}
\eff^{(i)}=\rho V_{b}\frac{\dee\overline{\you}}{\dee t} + 2\rho V_{b}\frac{\dee\Ohm}{\dee t}\times\overline{\ecks} - \rho \oint_{S_{b}}\ecks\times\dot{\gam}^{(i)} \dx
\label{eqFi}
\end{equation}
and

\begin{equation}
\eff^{(v)}= -\rho \oint_{S_{b}}\ecks\times\dot{\gam}^{(v)} \dx - \mu \int_{V_f}\ecks\times\nabla^{2}\ohm \dtwox .
\label{eqFv}
\end{equation}
\citet{eldredgeReconciliation} gives expressions for the force and moment on the body in terms of impulse changes over an infinitesimal time interval.  The formulae (\ref{eqFi}) and (\ref{eqFv}) are the continuous-time equivalents of his results for force.

Alternatively, the surface-traction representation gives,
\begin{equation}
\eff = \pee^{(i)} + \pee^{(v)} + \tee,
\label{eqFtraction}
\end{equation}
in which: $\pee^{(i)}$ is the contribution from the inviscid-pressure component, i.e.

\begin{equation}
\pee^{(i)}=\oint_{S_{b}}p^{(i)}\en \dx ;
\label{eqPoint}
\end{equation}
$\pee^{(v)}$ is similarly defined in terms of the viscous-pressure component $p^{(v)}$; and $\tee$ is the viscous-stress contribution.  This can be written as (cf.~Appendix \ref{appvisctrac})

\begin{equation}
\tee=\mu\oint_{S_b}\en\times\ohm \dx.
\label{eqtee}
\end{equation}
We thus expect that $\pee^{(i)}$ is equal to the expression for $\eff^{(i)}$ in (\ref{eqFi}), and $\pee^{(v)} + \tee$ to that for $\eff^{(v)}$ in (\ref{eqFv}). 

The link can be established with the help of an identity given by \citet[\S4.2]{saffman}:

\begin{equation}
\oint_{S_{b}}p\en \dx = -\oint_{S_{b}}\ecks\times(\en \times \nabla p) \dx .
\label{eqpid2d}
\end{equation}
Employing it in (\ref{eqPoint}), and invoking (\ref{eqgdi}), we obtain

\begin{equation}
\pee^{(i)} = - \rho \oint_{S_{b}}\ecks\times\dot{\gam}^{(i)} \dx + \rho\oint_{S_{b}}\ecks\times(\en\times\dot{\you}) \dx .
\label{eqFi2d}
\end{equation}
The second integral can be evaluated by applying Gauss's theorem to transpose it to the body interior and substituting the explicit formula (\ref{eqUdot}) for $\dot{\you}$.  The first step yields

\begin{equation}
\left[ \oint_{S_{b}}\ecks\times(\en\times\dot{\you})\dx \right]_{i} = \int_{V_{b}} \dot{U}_{i} \dtwox + \int_{V_{b}} x_{j} \left( \frac{\partial \dot{U}_{i}}{\partial x_{j}} - \frac{\partial \dot{U}_{j}}{\partial x_{i}} \right) \dtwox
\end{equation}
(with the summation convention in operation).  The second allows us to write $\partial \dot{U}_{i}/\partial x_{j} - \partial \dot{U}_{j}/\partial x_{i} = -2 \epsilon_{ijk} (\dee\Ohm/\dee t)_{k}$ ,
where $\epsilon_{ijk}$ is the Levi-Civita symbol (1 if $\left\{i,j,k\right\}$ is a cyclic permutation, $-1$ if it is anti-cyclic, 0 otherwise).  Thus

\begin{equation}
\oint_{S_{b}}\ecks\times(\en\times\dot{\you})\dx=V_{b}\frac{\dee\overline{\you}}{\dee t}+2V_{b}\frac{\dee\Ohm}{\dee t}\times\overline{\ecks} ,
\end{equation}
and (\ref{eqFi2d}) becomes

\begin{equation}
\pee^{(i)}=\rho V_{b}\frac{\dee\overline{\you}}{\dee t} + 2\rho V_{b}\frac{\dee\Ohm}{\dee t}\times\overline{\ecks} - \rho \oint_{S_{b}}\ecks\times\dot{\gam}^{(i)} \dx,
\end{equation}
which establishes the correspondence between $\pee^{(i)}$ and $\eff^{(i)}$.

For the viscous contribution, like \citet{eldredgeReconciliation}, we further develop the vorticity-moment expression (\ref{eqFv}).  The integrand in the second term can be written in the form

\begin{equation}
\left( \ecks\times\nabla^{2}\ohm \right)_{i} = \epsilon_{ijk} \left[ \frac{\partial}{\partial x_l} \left( x_j \frac{\partial\omega_{k}}{\partial x_l} \right) - \frac{\partial\omega_{k}}{\partial x_j} \right] .
\end{equation}
Thus, via Gauss's theorem and (\ref{eqtee}), (\ref{eqFv}) develops to

\begin{equation}
\eff^{(v)} = - \rho \oint_{S_b}\ecks\times\dot{\gam}^{(v)}\dx - \mu\oint_{S_{b}}\ecks\times\left(\en\cdot\nabla\ohm\right)\dx + \tee .
\label{eqFvdev}
\end{equation}
At this point, Eldredge follows \citet{lighthillBLs} in stating that the vorticity flux ${-\nu\en\cdot\nabla\ohm}$ maintains the no-slip condition by cancelling the inviscid vortex-sheet growth.  Specifically, this assertion implies $\nu\en\cdot\nabla\ohm=\dot{\gam}^{(i)}$ or, given that $\dot{\gam}^{(i)}+\dot{\gam}^{(v)}=0$,

\begin{equation}
\mu\en\cdot\nabla\ohm=-\rho\dot{\gam}^{(v)} .
\label{eqomflxlh}
\end{equation}
Hence, on this basis, the first two terms in (\ref{eqFvdev}) cancel; $\eff^{(v)}=\tee$, and the viscous-pressure contribution $\pee^{(v)}$ that appears in (\ref{eqFtraction}) is not recovered.

The problem lies in the Lighthill claim.  We show this by first noting the identity (in two dimensions)

\begin{equation}
\en\times\left(\nabla\times\ohm\right) = -\en\cdot\nabla\ohm ,
\label{eqennabla}
\end{equation}
which allows us to link $\en\cdot\nabla\ohm$ to the viscous component of the pressure decomposition.  Substituting (\ref{eqennabla}) into (\ref{eqgdv}) we obtain

\begin{equation}
\mu\en\cdot\nabla\ohm=-\rho\dot{\gam}^{(v)} + \en \times \nabla p^{(v)},
\label{eqomflux}
\end{equation}
differing from (\ref{eqomflxlh}) in the additional term $\en\times\nabla p^{(v)}$.  When (\ref{eqomflux}) is used in (\ref{eqFvdev}), this term yields, via (\ref{eqpid2d}), $\pee^{(v)}$, and thus

\begin{equation}
\eff^{(v)} = \pee^{(v)} + \tee,
\end{equation}
as expected.

The significance of the viscous-pressure component remains uncertain.  Given the scalings in the examples of Appendix \ref{appvpegs}, we would generally expect the inviscid pressure to exceed the viscous by a factor of order Reynolds number.  Hence it is tempting to state that $\pee^{(v)}$ is negligible in high-Reynolds-number flows, and that the error in Lighthill's claim is trivial.  The first point may well be true.  However, the second is not; in the three-dimensional case, as we shall see, it becomes a crucial obstacle to demonstrating the equivalence of the surface-traction and vorticity-moment formulations.

\subsection{Surface-traction/vorticity-moment comparison (three-dimensional)} \label{secstvm3d}
The formulae (\ref{eqdalphadt}) and (\ref{eqintVf}) remain valid in three dimensions, so we still have

\begin{equation}
\frac{\dee\boldsymbol{\alpha}}{\dee t} = -2V_{b}\frac{\dee\Ohm}{\dee t}\times\overline{\ecks} + \int_{V_f}\ecks\times\left(\frac{\partial\ohm}{\partial t}+\yoo\cdot\nabla\ohm\right)\dthreex
\label{eqdadt3d}
\end{equation}
for the rate of change of the first moment of vorticity.  However, the well-known additional vortex-stretching term now enters the vorticity equation, so

\begin{equation}
\frac{\partial\ohm}{\partial t}+\yoo\cdot\nabla\ohm = \ohm\cdot\nabla\yoo + \nu \nabla^{2} \ohm .
\end{equation}
The contribution of $\ohm\cdot\nabla\yoo$ to the integral in (\ref{eqdadt3d}) can be integrated by parts via the integrand manipulation

\begin{equation}
\left[ \ecks\times\left(\ohm\cdot\nabla\yoo\right) \right]_{i} = \epsilon_{ijk} \left[ \frac{\partial}{\partial x_l} \left( x_j \omega_l u_{k} \right) - \omega_{j} u_{k} \right]
\end{equation}
and Gauss's theorem, to give

\begin{equation}
\int_{V_{f}}\ecks\times(\ohm\cdot\nabla\yoo)\dthreex=\int_{V_{f}}\yoo\times\ohm \dthreex+\int_{S_{b}}(\ecks\times\yoo)\ohm\cdot\en \dtwox .
\end{equation}
Here the first term is given explicitly by (\ref{eqintVf}).  The second can be converted into an integral over the body volume with the aid of the boundary conditions $\yoo=\you$, $\ohm\cdot\en=2\Ohm\cdot\en$.  Remarkably, it turns out to cancel with the first, so (\ref{eqdadt3d}) simplifies to

\begin{equation}
\frac{\dee\boldsymbol{\alpha}}{\dee t} = -2V_{b}\frac{\dee\Ohm}{\dee t}\times\overline{\ecks} + \nu\int_{V_f}\ecks\times\nabla^{2} \ohm\dthreex ,
\end{equation}
as in two dimensions.  The expression for the force, however, differs subtly from (\ref{eqFWu2d}); it is given by \citep{wuTheory}

\begin{equation}
\eff=\rho V_{b}\frac{\dee\overline{\you}}{\dee t} - \frac{1}{2}\rho \frac{\dee\boldsymbol{\alpha}}{\dee t}.
\label{eqFWu3d}
\end{equation}
Hence, once more introducing the mutually cancelling vortex-sheet growth rates and writing $\eff = \eff^{(i)}+\eff^{(v)}$, we obtain

\begin{equation}
\eff^{(i)}=\rho V_{b}\frac{\dee\overline{\you}}{\dee t} + \rho V_{b}\frac{\dee\Ohm}{\dee t}\times\overline{\ecks} - \frac{1}{2}\rho \int_{S_{b}}\ecks\times\dot{\gam}^{(i)} \dtwox;
\end{equation}

\begin{equation}
\eff^{(v)}= -\frac{1}{2}\rho \int_{S_{b}}\ecks\times\dot{\gam}^{(v)} \dtwox - \frac{1}{2}\mu \int_{V_f}\ecks\times\nabla^{2}\ohm \dthreex .
\label{eqfv3x}
\end{equation}

Turning to the traction-based formulation, we again have $\eff=\pee^{(i)}+\pee^{(v)}+\tee$.  However, the identity (\ref{eqpid2d}) gains a factor 1/2 in three dimensions, so now

\begin{equation}
\int_{S_{b}}p\en \dtwox = -\frac{1}{2}\int_{S_{b}}\ecks\times(\en \times \nabla p) \dtwox.
\label{eqpid3d}
\end{equation}
Following the same route as in two dimensions, (\ref{eqgdi}) is used to express $\en\times\nabla p^{(i)}$ in terms of $\dot{\gam}^{(i)}$ and $\dot{\you}$, and the integral involving the latter is converted from the body surface to the body volume.  The manipulation differs slightly, but the upshot is --- as before --- that $\pee^{(i)}=\eff^{(i)}$.  Again, we find that the force associated with the inviscid-pressure component $p^{(i)}$ is consistent with that derived from the inviscid part of the vorticity-moment/impulse formulation.

The reconciliation of the viscous-force expressions, however, does not follow the two-dimensional case so closely.  The first step is the same, i.e. manipulating the second integral in (\ref{eqfv3x}) into body-surface form, but (by virtue of the difference between the three- and two-dimensional expressions for the force, (\ref{eqFWu3d}) and (\ref{eqFWu2d})) the starting point differs from its two-dimensional counterpart, (\ref{eqFv}), by a factor of two.  Thus we obtain

\begin{equation}
\eff^{(v)} = - \frac{1}{2}\rho \int_{S_b}\ecks\times\dot{\gam}^{(v)}\dtwox - \frac{1}{2}\mu\int_{S_{b}}\ecks\times\left(\en\cdot\nabla\ohm\right)\dtwox + \frac{1}{2}\tee, 
\label{eqFv3d2}
\end{equation}
exactly half the previous expression, (\ref{eqFvdev}).  Recalling that Lighthill's assertion implies cancellation of the first two terms, we see that now it not only fails to account for the contribution of $p^{(v)}$, but also for half of the resultant viscous-stress force.

The solution is again to employ the exact expression for $\en\cdot\nabla\ohm$ implied by the Navier-Stokes equation and the pressure-field decomposition.  In three dimensions, (\ref{eqennabla}) is replaced by 

\begin{equation}
\left[\en\times\left(\nabla\times\ohm\right)\right]_{i} = -\en\cdot\nabla\omega_{i} + n_{j}\frac{\partial\omega_{j}}{\partial x_{i}},
\end{equation}
so (\ref{eqgdv}) now yields

\begin{equation}
\mu\en\cdot\nabla\ohm=-\rho\dot{\gam}^{(v)} + \en \times \nabla p^{(v)} + \mu n_{j}\nabla\omega_{j} .
\label{eqomflux3d}
\end{equation}
Comparing against (\ref{eqomflux}), we observe that the shift to three dimensions has introduced the additional final term, and it is this that supplies the missing shear contribution (see Appendix \ref{appmissvs}).  Thus (\ref{eqFv3d2}) becomes

\begin{equation}
\eff^{(v)} = - \frac{1}{2} \int_{S_b}\ecks\times(\en \times \nabla p^{(v)})\dtwox + \tee ,
\end{equation}
and, by virtue of (\ref{eqpid3d}), $\eff^{(v)}=\pee^{(v)}+\tee$.

This completes the reconciliation of the surface-traction and vorticity-moment formulations in three dimensions.  However, the mathematical analysis does not explain \emph{why} the intuitively attractive viewpoint whereby the viscous vorticity flux exactly cancels the inviscid growth of slip is wrong.  A conceptual interpretation of this point is given in Appendix \ref{appvfandslip}.

\subsection{Summary}
This part of the paper has addressed a discrepancy between the force components arising from the viscous/inviscid pressure-field decomposition of \S\ref{secviscinv} and those identified by \citet{eldredgeReconciliation} on the basis of the vorticity-moment formulation.  We have found that the inviscid and viscous components of `impulse' defined by Eldredge remain valid, but that there is an omission in his analysis linking viscous impulse to the viscous-stress traction force. The omission arises from Lighthill's claim that $-\nu\en\cdot\nabla\ohm$ is equal and opposite to the rate of growth of surface slip implied by the inviscid flow component.  The correct expression for $\nu\en\cdot\nabla\ohm$, which can be obtained from the governing equations, recovers the complete traction contribution due to viscosity.

\section{The accelerative forces and added mass} \label{secimp4am}
In this section, we continue the discussion on the relative merits of the potential/circulatory and convective/accelerative decompositions that was begun in \S\ref{secinvcomp}.  Specifically, we compare the force resultants arising from the `accelerative' pressure component with their potential-flow counterparts (`added mass').  It might be argued that the familiarity of the latter makes the potential/circulatory decomposition preferable, even if it cannot be interpreted in terms of independent physical processes.  However, the comparison will demonstrate that this advantage is offset by the complexity of the general added-mass formulae, and that these formulae are actually further from an intuitive representation of inertia than the expressions for the accelerative force resultants.

At this point, it must be acknowledged that a universally accepted definition of added mass for general body motions does not currently exist.  Indeed, the term is often used loosely.  However, the apparent theoretical consensus \citep[cf., for example,][]{howeOnForce, eldredgeReconciliation} is to identify it directly with the potential-flow force resultants, and this convention will be followed here.  Given this choice, the added-mass force and moment are described in \S\ref{secamfrs}.  Then, in \S\ref{secaccfrs}, the accelerative force resultants are discussed.  These quantities can be expressed in terms of the classical inertia tensors that arise in the added-mass representation, allowing a direct comparison to be made.  The upshot is summarised in \S\ref{secamsum}.

\subsection{Force resultants in potential flow} \label{secamfrs}
General expressions for the force resultants on a body in potential flow have long been established \citep[the classic reference is][Ch.~VI]{lamb}.  However, no single source sets out the derivation in a form that is both transparent to present-day readers and suitable for our current purposes.  Hence it is given here in full.

Consider the external force, $\eff^{(ep)}$, that must be applied to the body to achieve its specified motion.  Direct evaluation of $\eff^{(ep)}$ via the integral momentum equation is complicated by the indeterminacy of the contribution from the fluid velocity, $\int_{V_{f}}\yoo \,\dee V$ \citep[cf., for example,][\S6.4]{batchelorFluidDyn}.  However, it can be expressed in terms of the impulse, $\mathbf{I}$, that must be applied to the body to generate the current velocity field instantaneously from rest.  For potential flow, this quantity is a well-defined property of the system, because the fluid velocity $\yoo\,(=\nabla\phi)$ is uniquely specified by the body velocity \citep[\S\S2.9, 2.10]{batchelorFluidDyn}.  The link to the external force on the body is made by considering three steps: (i) starting from the stationary body/fluid system, apply the impulse $\mathbf{I}$ required to produce the instantaneous flow; (ii) allow the system to evolve under the influence of the external force $\eff^{(ep)}$ for a short time interval $\delta t$; (iii) apply the impulse $-(\mathbf{I} + \delta\mathbf{I})$ needed to bring the system back to rest.  There is no system momentum at either the beginning or end of this process, so Newton's second law can be applied without ambiguity: the net external impulse that has been applied, $\mathbf{I} + \eff^{(ep)}\delta t -(\mathbf{I} + \delta\mathbf{I})$, is zero.  Thus, in the limit,

\begin{equation}
\eff^{(ep)} = \frac{\dee \mathbf{I}}{\dee t} .
\label{eqFexI}
\end{equation}

For this observation to be useful, we require an expression for the impulse.  This is found by considering the quantities acting on the body during the impulsive flow generation, namely $\mathbf{I}$ and the fluid pressure field.  The latter can be found from the unsteady Bernoulli equation, which is \citep[\S6.2]{batchelorFluidDyn}

\begin{equation}
\rho \frac{\partial\phi}{\partial t} + p + \frac{1}{2} \rho \left| \yoo \right|^{2} = \mathrm{constant},
\end{equation}
to within an arbitrary (and irrelevant) function of time.  The right-hand side and the final term on the left are finite during the infinitesimally small duration of the impulse application, so integrating over this period (and noting that $\phi = 0$ initially) shows that the impulsive pressure field is $-\rho\phi$, with $\phi$ the potential of the fluid velocity field that has been established.  Hence, on consideration of the body momentum change alone,

\begin{equation}
\mathbf{I} + \int_{S_{b}} (-\rho\phi) \en \,\dee S = \rho_{b}V_{b}\overline{\you} ,
\label{eqimp}
\end{equation}
with $\rho_{b}$ the body's density.  (For simplicity, $\rho_{b}$ is taken to be constant, so the body's centre of volume and centre of mass coincide.  The general case complicates the algebra without affecting the features of interest to us: the force resultants associated with the fluid.)

In principle, this completes the solution for the external force applied to the body; given the body motion, the velocity potential $\phi$ can be calculated as a function of time, with $\mathbf{I}(t)$ following from (\ref{eqimp}) and $\eff^{(ep)}$ from (\ref{eqFexI}).  However, a more illuminating formulation arises from expressing $\phi$ in terms of component functions associated with unit motions in each of the body's degrees of freedom.  To do so, we require a set of axes fixed in the body.  The associated (orthogonal) unit vectors will be denoted $\left( \eee_{1}, \eee_{2}, \eee_{3} \right)$.  Then

\begin{equation}
\phi = \sum_{j}\phi_{j}^{(t)}\eee_{j}\cdot\overline{\you} + \sum_{j}\phi_{j}^{(r)}\eee_{j}\cdot\Ohm ,
\label{eqphicomps}
\end{equation}
where $\phi_{j}^{(t)}$ is the velocity potential associated with unit body velocity in the $\eee_{j}$ direction, and $\phi_{j}^{(r)}$ is that associated with unit angular velocity about the $\eee_{j}$ axis.  These quantities are uniquely determined by the requirements that they satisfy Laplace's equation, and that the no-penetration conditions

\begin{equation}
\en\cdot\nabla\phi_{j}^{(t)} = \en\cdot\eee_{j} , \qquad \en\cdot\nabla\phi_{j}^{(r)} = \en\cdot\left[\eee_{j}\times\left(\ecks-\overline{\ecks}\right)\right]
\label{eqphibcs}
\end{equation}
apply on the body surface.  (In two dimensions, there must also be no circulation associated with either.)

Employing (\ref{eqphicomps}) in (\ref{eqimp}) allows it to be written as

\begin{equation}
\mathbf{I} = \sum_{i} I_{i} \eee_{i} ,
\label{eqimpba}
\end{equation}
with components

\begin{equation}
I_{i} = \rho_{b}V_{b}\left(\eee_{i}\cdot\overline{\you}\right)  + \sum_{j} M_{ij} \left(\eee_{j}\cdot\overline{\you}\right) + \sum_{j} N_{ij} \left(\eee_{j}\cdot\Ohm\right) ,
\label{eqimpcomp}
\end{equation}
in which

\begin{equation}
M_{ij} = \rho \eee_{i} \cdot \int_{S_{b}} \phi_{j}^{(t)} \en \,\dee S
\end{equation}
and

\begin{equation}
N_{ij} = \rho \eee_{i} \cdot \int_{S_{b}} \phi_{j}^{(r)} \en \,\dee S .
\label{eqnij}
\end{equation}
As they define the fluid part of the impulse in terms of $\overline{\you}$ and $\Ohm$, these tensor quantities can be viewed as inertia coefficients.  They depend only on the shape of the body.

In the form expressed by (\ref{eqFexI}), (\ref{eqimpba}), and (\ref{eqimpcomp}), the effect of the fluid is not obviously to add apparent mass.  Indeed, in the sense of rigid-body dynamics (as exemplified by the first term in (\ref{eqimpcomp})), it only does so for special cases.  A sphere of radius $a$, for example, has $M_{ij}$ equal to zero for $i \neq j$, and to $\frac{2}{3}\pi\rho a^{3}$ for $i = j$.  It also has zero contribution from its angular velocity, so (\ref{eqimpba}) becomes

\begin{equation}
\mathbf{I} = \left( \rho_{b}V_{b} + \frac{2}{3}\pi\rho a^{3} \right) \overline{\you} ,
\end{equation}
which has the expected form.  More generally, \citet{{limachergenderiv}} regard (\ref{eqimpcomp}) as admitting a generalised-inertia analogy for any body if $\Ohm = 0$.  From this viewpoint, inertia is characterised by a linear dependence of $\mathbf{I}$ on $\overline{\you}$, irrespective of whether the two are parallel.

A corresponding derivation, in terms of moment of momentum, yields the external moment that must be applied to produce the specified body motion.  Thus we define $\mathbf{H}$ as the torque impulse required (in conjunction with $\mathbf{I}$ acting at the body's centre of volume) to generate the current body and fluid velocities.  The link to the external moment, $\mathbf{Q}^{(ep)}$, is established exactly as previously; here, for the three-step process that starts and finishes with the system stationary, the net moment of applied impulses is zero.  This condition yields

\begin{equation}
\mathbf{Q}^{(ep)} = \overline{\you}\times\mathbf{I} + \frac{\dee \mathbf{H}}{\dee t} .
\label{eqqam}
\end{equation}
Meanwhile, consideration of the moment-of-momentum change of the body alone during the impulsive start gives

\begin{equation}
\eee_{i} \cdot \left\{ \mathbf{H} + \int_{S_{b}} (-\rho\phi) \left( \ecks-\overline{\ecks} \right) \times \en \,\dee S \right\} = \sum_{j} J_{ij} \left(\eee_{j}\cdot\Ohm\right) ,
\end{equation}
where the $J_{ij}$ are the values of the body's moment-of-inertia tensor in the body-axis system.  As before, $\phi$ is written in the form (\ref{eqphicomps}), leading to

\begin{equation}
\eee_{i} \cdot \mathbf{H} = \sum_{j} J_{ij} \left(\eee_{j}\cdot\Ohm\right) + \sum_{j} N_{ji} \left(\eee_{j}\cdot\overline{\you}\right) + \sum_{j} R_{ij} \left(\eee_{j}\cdot\Ohm\right) ,
\label{eqmofi}
\end{equation}
where the third inertia tensor, $R_{ij}$, is given by

\begin{equation}
R_{ij} = \rho \eee_{i} \cdot \int_{S_{b}} \phi_{j}^{(r)} \left(\ecks-\overline{\ecks}\right) \times \en \,\dee S .
\label{eqrij}
\end{equation}
(The term $N_{ji}$ appears because

\begin{equation}
\eee_{i} \cdot \int_{S_{b}} \phi_{j}^{(t)} \left(\ecks-\overline{\ecks}\right) \times \en \,\dee S = \eee_{j} \cdot \int_{S_{b}} \phi_{i}^{(r)} \en \,\dee S ,
\end{equation}
as can be shown with the aid of (\ref{eqphibcs}), Gauss's theorem, and the identity $\eee_{i} \cdot \left(\ecks-\overline{\ecks}\right) \times \en = \en \cdot \eee_{i} \times \left(\ecks-\overline{\ecks}\right)$.)

\subsection{The accelerative force resultants} \label{secaccfrs}
We now consider the counterparts of the potential-flow external force resultants for the accelerative pressure component, $p^{(a)}$.  These are the force and moment that must be applied to the stationary body (in stationary fluid) to give it linear acceleration $\dee\overline{\you}/\dee t$ and angular acceleration $\dee\Ohm/\dee t$.  Denoting them as $\eff^{(ea)}$ and $\mathbf{Q}^{(ea)}$, we have, from the equations of motion for the body,

\begin{equation}
\eff^{(ea)} + \int_{S_{b}}p^{(a)}\en \,\dee S = \rho_{b} V_{b} \frac{\dee\overline\you}{\dee t}
\label{eqPaint}
\end{equation}
and 

\begin{equation}
\eee_{i} \cdot \left\{ \mathbf{Q}^{(ea)} + \int_{S_{b}}p^{(a)} (\ecks - \overline{\ecks})\times\en \,\dee S \right\} = \sum_{j}J_{ij}\left(\eee_{j}\cdot\frac{\dee\Ohm}{\dee t}\right) .
\label{eqQaint}
\end{equation}
The accelerative pressure is specified by (\ref{eqpa}) and (\ref{eqpabc}), which show that it satisfies Laplace's equation with a combination of uniform translation and uniform rotation boundary conditions.  Hence it too can be expressed in terms of the unit potentials of \S\ref{secamfrs}, as

\begin{equation}
p^{(a)} = -\rho\sum_{j}\phi_{j}^{(t)}\left(\eee_{j}\cdot\frac{\dee\overline\you}{\dee t}\right) - \rho\sum_{j}\phi_{j}^{(r)}\left(\eee_{j}\cdot\frac{\dee\Ohm}{\dee t}\right) .
\end{equation}
On substituting this representation into (\ref{eqPaint}) and (\ref{eqQaint}), we obtain

\begin{equation}
\eee_{i} \cdot \eff^{(ea)} = \rho_{b} V_{b} \left(\eee_{i}\cdot\frac{\dee\overline\you}{\dee t}\right) + \sum_{j}M_{ij}\left(\eee_{j}\cdot\frac{\dee\overline\you}{\dee t}\right) + \sum_{j}N_{ij}\left(\eee_{j}\cdot\frac{\dee\Ohm}{\dee t}\right) .
\label{eqfacc}
\end{equation}
and

\begin{equation}
\eee_{i} \cdot \mathbf{Q}^{(ea)} = \sum_{j}J_{ij}\left(\eee_{j}\cdot\frac{\dee\Ohm}{\dee t}\right) + \sum_{j}N_{ji}\left(\eee_{j}\cdot\frac{\dee\overline\you}{\dee t}\right) + \sum_{j}R_{ij}\left(\eee_{j}\cdot\frac{\dee\Ohm}{\dee t}\right) .
\label{eqqacc}
\end{equation}

The comparison with $\eff^{(ep)}$ and $\mathbf{Q}^{(ep)}$ requires the time derivatives in (\ref{eqFexI}) and (\ref{eqqam}) to be implemented explicitly.  Here the rotation of the body-fixed axes must be taken into account, via the standard result $\dee\eee_{i}/\dee t = \Ohm\times\eee_{i}$.  This means that additional terms beyond those in (\ref{eqfacc}) and (\ref{eqqacc}) appear.  Specifically, from (\ref{eqFexI}) in conjunction with (\ref{eqimpba}) and (\ref{eqimpcomp}), we find

\begin{equation}
\eee_{i} \cdot \eff^{(ep)} = \eee_{i} \cdot \eff^{(ea)} + 
\eee_{i} \cdot \left[\Ohm\times\left(\mathbf{I}-\rho_{b}V_{b}\overline{\you}\right)\right] + \sum_{j} \left(\Ohm\times\eee_{j}\right) \cdot \left( M_{ij}\overline\you \right) .
\label{eqFecfFa}
\end{equation}
Similarly, (\ref{eqqam}) and (\ref{eqmofi}) lead to

\begin{equation}
\eee_{i} \cdot \mathbf{Q}^{(ep)} = \eee_{i} \cdot \mathbf{Q}^{(ea)} + \eee_{i} \cdot \left( \overline{\you}\times\mathbf{I} \right) + \eee_{i} \cdot \left(\Ohm\times\mathbf{H}\right) + \sum_{j} \left(\Ohm\times\eee_{j}\right) \cdot \left( N_{ji}\overline\you \right) .
\label{eqQecfQa}
\end{equation}

That there are differences between the accelerative and added-mass expressions is hardly surprising, given their respective definitions.  More interesting is that the accelerative force resultants are intuitively closer to an interpretation as additional inertia than their potential-flow counterparts.  Some of the extra terms in (\ref{eqFecfFa}) and (\ref{eqQecfQa}) can be interpreted as centripetal and gyroscopic effects, but the others have no counterparts in the dynamics of a rigid body.  Admittedly, even (\ref{eqfacc}) and (\ref{eqqacc}) demand an extension to the concept of inertia; the angular acceleration contributes to the fluid force via $N_{ij}$, and the linear acceleration to the fluid moment likewise.  Also, the mass-like effect of the fluid appears as a tensor, $M_{ij}$, rather than a scalar.  Nonetheless, these features seem reasonably natural, unlike the further elements that appear in the potential-flow expressions.  Thus it can be argued that the accelerative force resultants are, in fact, more appropriate as representations of effective additional inertia than those conventionally termed `added mass'.

This said, there is one special case where the two coincide: the force on the body in purely translational motion.  Now $\Ohm = 0$, and the components $\eee_{i}\cdot\eff^{(ep)}$ and $\eee_{i}\cdot\eff^{(ea)}$ both become

\begin{equation}
\rho_{b} V_{b} \left(\eee_{i}\cdot\frac{\dee\overline\you}{\dee t}\right) + \sum_{j}M_{ij}\left(\eee_{j}\cdot\frac{\dee\overline\you}{\dee t}\right) .
\end{equation}
The correspondence is simply a reflection of the d'Alembert paradox, which implies that the potential-flow force in this case arises solely from the body acceleration.  \citep[For an explicit demonstration, see][\S6.4, where the force on a translating body in potential flow is derived via the unsteady Bernoulli equation.]{batchelorFluidDyn}  Note that it does not extend to the moment, which is non-zero in the steady potential flow around a body without any specific geometrical symmetries.

\subsection{Summary} \label{secamsum}
In this part of the paper, we have compared the classical, added-mass, force resultants with the contributions from the `accelerative' pressure component identified in \S\ref{secinvcomp}, and demonstrated that the latter are more easily interpreted as inertial than the explicit expressions for added-mass force and moment.  There is, however, one exception: when the body motion is purely translational, the force due to the accelerative pressure corresponds exactly to its added-mass counterpart.  Hence the analysis presented here is consistent with the theoretical arguments of \citet[appendix]{leonardAM}, and the experimental findings of \citet{limacherpiv,corkeryAM}.

\section{Conclusions}
This paper has presented a novel decomposition of the force resultants on a body moving in an unbounded, incompressible fluid.  It rests on a partition of the pressure field into `viscous' and `inviscid' components, with the latter consisting of separate `convective' and `accelerative' contributions.  The inviscid pressure is defined by (\ref{eqd2pi}) and (\ref{eqpibc}), and is that which would arise from the same body motion in a viscosity-free fluid with the same instantaneous velocity field.  The viscous pressure obeys (\ref{eqd2pv}) and (\ref{eqpvbc}), and enters because the inviscid component alone does not satisfy the requisite boundary condition at the body surface.  Together with the viscous stresses, it gives rise to a force and moment exclusively due to viscosity.

The other, inviscid, force resultants come purely from pressure traction, with convective and accelerative parts specified via (\ref{eqpc})--(\ref{eqpabc}).  The convective resultants would be observed if the body's centroid and angular velocities at the given instant were unchanging, while the accelerative resultants correspond to the body accelerating from rest in quiescent fluid.  The convective and accelerative quantities are mutually independent, and can be identified without ambiguity for any given case.  Likewise, they are both independent of the viscous resultants.

In the context of previous work, two questions arise from the new formulation.  The first is the absence of a viscous-pressure contribution in a viscous/inviscid decomposition based on the vorticity-moment expressions for the force resultants \citep{eldredgeReconciliation}.  The source of this discrepancy has been identified as an error in the \citet{lighthillBLs} claim that the viscous vorticity flux annihilates the vortex sheet that would develop at the body surface if the fluid were inviscid.  The second issue is the relation of the accelerative force resultants to their `added-mass' counterparts.  These arise from the well-known split of the fluid velocity into an irrotational component consistent with the body's motion and a rotational remainder; they are the resultants associated with the irrotational part.  It has been argued here that, in a general flow, they do not fully represent the force resultants due to the body motion.  Thus, for this reason alone, the accelerative resultants should instead be the focus of attention.  Moreover, direct comparison of the respective general formulae shows that the accelerative resultants are in fact more naturally interpretable as additional inertia than the so-called added-mass contributions.

On this basis, the current decomposition is recommended to experimental and numerical fluid-mechanics practitioners seeking to understand and/or model the force resultants on moving bodies.  As a minimum, given an independent assessment of overall force and moment, one can obtain the accelerative resultants from (\ref{eqfacc}) and (\ref{eqqacc}), and the remainder is then the combined convective and viscous contribution.  Separate calculation of the convective part is also possible; as with the vorticity-moment formulation, only the velocity field and its spatial gradients are required.  (A complicating factor in practice will be the need to specify a finite domain with a boundary condition for the convective pressure on its outer surface, but, in mitigation, it will be unnecessary to perform a differentiation with respect to time.)  Direct evaluation of the viscous part, however, demands \emph{vorticity} gradients, and may thus be out of reach for present-day experimental datasets.\\

This work was undertaken without external funding.  The implicit financial support provided by the University of Cambridge is gratefully acknowledged.  Thanks are also due to Ignacio Andreu Angulo, Holger Babinsky, Simon Corkery, Jeff Eldredge, Pascal Gehlert, Anya Jones and Jie Li for their comments on draft versions of the paper.  In particular, Appendices \ref{appvpegs} and \ref{appvfandslip} arose directly from questions posed by Eldredge and Jones respectively.  Finally, the thoughtful and constructive criticisms put forward by the anonymous reviewers were extremely helpful.

\appendix
\section{Examples of the viscous-pressure component} \label{appvpegs}
As a `viscous' pressure field is not an intuitively natural concept, it is instructive to see how it appears in exact solutions to the Navier-Stokes equation.  Unfortunately, there are (to the author's knowledge) no useful solutions for closed bodies; those that are relevant involve surfaces of infinite extent.  This means that some artificiality must be accepted.  Subject to this caveat, the appendix considers Poiseuille flow, stagnation flow, and `von K\'arm\'an's viscous pump'.  In all three cases, the most straightforward form of the boundary condition (\ref{eqpvbc}) is the alternative that follows from (\ref{eqd2ueqcu}), i.e.

\begin{equation}
\en\cdot\nabla p^{(v)} = \mu\en\cdot\nabla^{2}\yoo .
\label{eqpvbc2}
\end{equation}

\subsection{Poiseuille flow}
The class of parallel flows, in which $\yoo\cdot\nabla\yoo = 0$, gives rise to numerous exact solutions of the Navier-Stokes equation \citep{wangExactNS}.  One of the simplest, flow between parallel plates, is sufficient to illustrate the r\^ole of the viscous-pressure component in these cases.  The solution is standard \citep[cf., for example,][\S4.2]{batchelorFluidDyn}: the velocity $\yoo = (u,0,0)$ between surfaces at $y = \pm h$ is given by

\begin{equation}
u = \frac{1}{2\mu} \left( -\frac{\dee p}{\dee x} \right) (h^{2}-y^{2}) ,
\end{equation}
with $\dee p/\dee x$ constant.

From (\ref{eqd2pi}) and (\ref{eqd2pv}), both $p^{(i)}$ and $p^{(v)}$ must satisfy Laplace's equation, while (\ref{eqpibc}) and (\ref{eqpvbc2}) show that

\begin{equation}
\en\cdot\nabla p^{(i)} = \en\cdot\nabla p^{(v)} = 0
\end{equation}
on the plate surfaces.  The decomposition thus rests on the boundary conditions on planes of constant $x$, at $0$ and $X$ say.  These lie in the fluid, so the body-surface form (\ref{eqpbc}) is no longer appropriate.  Specifically, reversing a step taken in the original derivation, the body acceleration $\dot{\you}$ must be replaced by the material derivative of the fluid velocity.  This affects (\ref{eqpibc}) for the inviscid component, but not (\ref{eqpvbc2}) for the viscous.  Now, with $\en$ equal to the unit $x$-direction vector $\mathbf{e}_{x}$ on the right-hand boundary, and $-\mathbf{e}_{x}$ on the left, these equations become

\begin{equation}
\frac{\partial}{\partial x}p^{(i)} = 0 ,
\end{equation}
and

\begin{equation}
\frac{\partial}{\partial x}p^{(v)} = \frac{\dee p}{\dee x} .
\end{equation}
Hence, by inspection, $p^{(i)} = 0$ and $p^{(v)} = p$; the viscous component is the sole contributor to the pressure field.  This is consistent with the physical interpretation of the inviscid pressure (\S\ref{secviscinv}), because no pressure gradient is needed to maintain the steady parallel flow of an inviscid fluid.

\subsection{Stagnation flow}
As already noted, the contribution of the viscous-pressure component in stagnation flow is evident in the results for total pressure given by \citet{issaTotp} and \citet{williamsTotp}.  Here we show how it arises from the defining equations of \S\ref{secviscinv}.

For an ideal fluid, the two-dimensional stagnation flow approaching the plane $y = 0$ in the negative-$y$ direction has $\yoo = (Bx,-By)$, with $B$ constant.  The corresponding viscous flow has a classical exact solution; this description follows the presentation given by \citet[\S3-8.1]{whiteViscFlo}.

The velocity components $(u,v)$ are

\begin{equation}
u = BxF^{\prime}(\eta) , \qquad v = -(B\nu)^{1/2}F(\eta) ,
\label{eqspuv}
\end{equation}
where $F(\eta)$ can be found numerically.  Its (dimensionless) argument, $\eta$, is given by

\begin{equation}
\eta = (B/\nu)^{1/2}y .
\end{equation}
The function $F$ has zero value and gradient at $\eta = 0$ (so that the no-slip condition is satisfied), and $F^{\prime}(\eta) \to 1$ as $\eta \to \infty$ (so that the potential-flow solution for $u$ is regained).  Note that, as the velocity increases without limit away from the origin, the solution is unphysical in a global sense; it can only be regarded as a local representation of a real stagnation flow.

The viscous-pressure component satisfies Laplace's equation and, via (\ref{eqpvbc2}), 

\begin{equation}
\left[ \frac{\partial}{\partial y}p^{(v)} \right]_{y=0} = -\rho B^{3/2} \nu^{1/2} F^{\prime\prime}(0) .
\end{equation}
Hence, to within an arbitrary constant, it is given by

\begin{equation}
p^{(v)} = -\rho B \nu F^{\prime\prime}(0) \eta .
\label{eqsppv}
\end{equation}
The overall pressure can be derived in terms of $F$ and its derivatives by substituting the velocity components (\ref{eqspuv}) into the Navier-Stokes equation (\ref{eqns}) and integrating.  However, it is more instructive to consider the gradient, which is the entity relevant to the flow dynamics.  In terms of the dimensionless pressure $\tilde{p} = p/\frac{1}{2} \rho B \nu$, we have

\begin{equation}
\frac{\partial \tilde{p}}{\partial \eta} = -2 \left[ F(\eta)F^{\prime}(\eta) + F^{\prime\prime}(\eta) \right] .
\end{equation}
The viscous component of this quantity is, by (\ref{eqsppv}), $-2 F^{\prime\prime}(0)$, and the inviscid part is the remainder.  Given the boundary conditions on $F$, it has value zero at $\eta = 0$, in agreement with (\ref{eqpibc}).

\begin{figure}
\begin{center}
\includegraphics[scale=0.6]{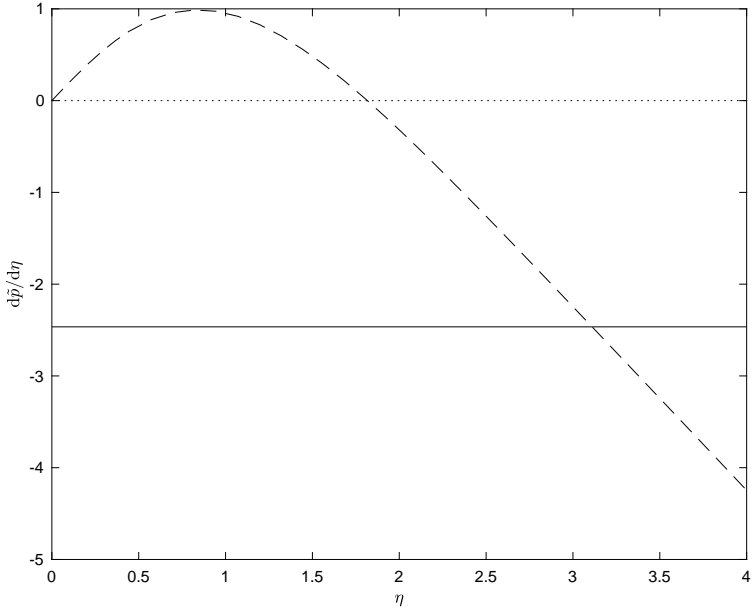}
\caption{{\label {spdpdy}} Viscous (---) and inviscid (-- --) components of the dimensionless $y$-direction pressure gradient in the Navier-Stokes solution for stagnation-point flow.}
\end{center}
\end{figure}

Figure \ref{spdpdy} shows the viscous and inviscid components of $\partial \tilde{p}/\partial \eta$.  The constant viscous part seems contradictory, given that viscous effects in this flow become unimportant away from the wall.  However, for large $\eta$, the inviscid part grows like $\eta$, and so becomes dominant.

\subsection{Von K\'arm\'an's viscous pump}
The flow above a surface in the plane $z = 0$, spinning with angular velocity $\Omega$ about the $z$ axis, has an exact solution originally given by von K\'arm\'an \citep[\S3-8.2]{whiteViscFlo}.  The fluid above the surface rotates by virtue of the no-slip condition and viscous shear.  There is no radial pressure gradient, so the fluid also centrifuges outwards.  Volume conservation requires a matching inwards flow, which is parallel to the $z$ axis.  Like the previous example, this solution is unphysical in a global sense; somewhere `outside' it the radial outflow must be recycled to provide the axial inflow.  However, the velocity and pressure remain finite as $z \to \infty$.

In rectangular polar coordinates $(r,\theta,z)$, the velocity field is given by

\begin{equation}
u_{r} = r \Omega F(\zeta) , \qquad u_{\theta} = r \Omega G(\zeta) , \qquad u_{z} = (\Omega\nu)^{1/2} H(\zeta) ,
\end{equation}
where $\zeta$ is a dimensionless $z$ coordinate:

\begin{equation}
\zeta = (\Omega/\nu)^{1/2} z .
\end{equation}
The functions $F(\zeta)$, $G(\zeta)$ and $H(\zeta)$ must be determined numerically.  A fourth, uncoupled, function $P(\zeta)$ describes the pressure, via

\begin{equation}
p = \rho \Omega \nu P(\zeta) .
\end{equation}
The boundary conditions on the functions are: $F(0) = H(0) = P(0) = 0$; $G(0) = 1$; $F(\zeta),\,G(\zeta) \to 0$ as $\zeta \to \infty$.  Note that these imply a non-zero pressure at infinity; its value could be subtracted from the pressure field without affecting the flow dynamics.

As in the stagnation-flow example, the normal gradient of the inviscid-pressure field, $p^{(i)}$, is zero at the surface (cf.~(\ref{eqpibc})), so $\dee p/\dee z$ there is entirely associated with the viscous pressure.  On the basis of this observation, and the relation

\begin{equation}
P^{\prime}(\zeta) = 2F(\zeta)H(\zeta) - 2F^{\prime}(\zeta)
\label{eqpdash}
\end{equation}
(one of the governing equations),

\begin{equation}
\left[ \frac{\dee}{\dee z}p^{(v)} \right]_{z=0} = -2\rho \Omega^{3/2} \nu^{1/2} F^{\prime}(0) .
\end{equation}
Once again, the viscous-pressure field is linear in distance from the surface.  Defining $\tilde{p} = p/\frac{1}{2} \rho \Omega \nu$ in this case, we have

\begin{equation}
\tilde{p}^{(v)} = -4\zeta F^{\prime}(0) .
\label{eqpvofzet}
\end{equation}
The overall dimensionless pressure is

\begin{equation}
\tilde{p} = 2P(\zeta) ,
\label{eqptil}
\end{equation}
and $\tilde{p}^{(i)}$ = $\tilde{p} - \tilde{p}^{(v)}$.

\begin{figure}
\begin{center}
\includegraphics[scale=0.6]{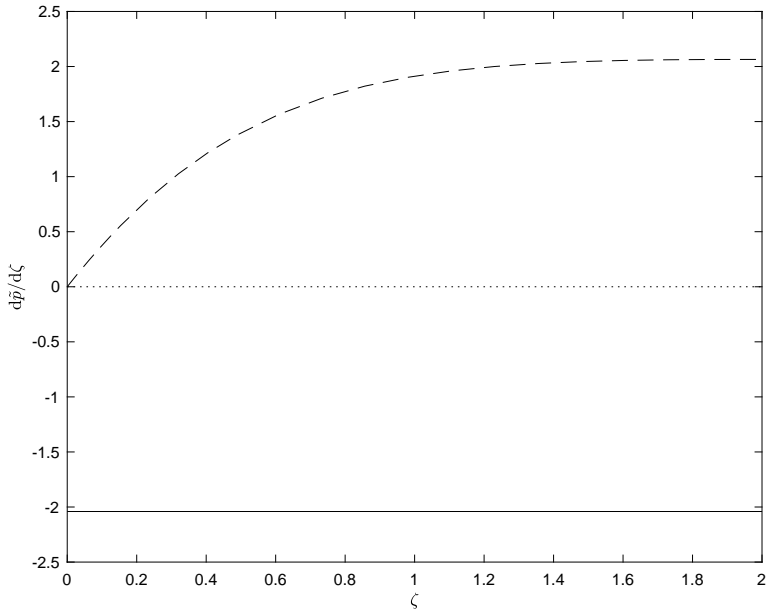}
\caption{{\label {vkvpdpdy}} Viscous (---) and inviscid (-- --) components of the dimensionless $z$-direction pressure gradient in the Navier-Stokes solution for flow over a spinning surface.}
\end{center}
\end{figure}

As with the stagnation flow, we consider the gradients of pressure components.  From (\ref{eqpvofzet}), we have $\dee \tilde{p}^{(v)}/\dee \zeta = -4 F^{\prime}(0)$, while $\dee \tilde{p}/\dee \zeta = 2 P^{\prime}(\zeta)$, with $P^{\prime}(\zeta)$ given by (\ref{eqpdash}).  Finally, $\dee \tilde{p}^{(i)}/\dee \zeta = \dee \tilde{p}/\dee \zeta - \dee \tilde{p}^{(v)}/\dee \zeta$.  The results are plotted in Fig.~\ref{vkvpdpdy}.  In this case, the inviscid component does not eventually dominate; instead it grows until it exactly cancels its viscous counterpart.

Given the entirely viscous nature of the current example, the lesser importance of the inviscid component (compared to the stagnation flow) is not surprising.  However, the continuous growth of both pressure components in a case where their resultant tends to a constant value might call the physical legitimacy of the viscous/inviscid decomposition into question.  The response is that this behaviour is an artefact associated with the assumed problem geometry.  For a realistic, closed, body shape, the viscous component decays to zero (or an arbitrary constant) at infinity.  This can be shown as follows.

From (\ref{eqd2pv}) and (\ref{eqpvbc2}), it is evident that finding the behaviour of $p^{(v)}$ far away from the body is directly analogous to the same problem for the velocity potential, $\phi$, in irrotational flow.  The analysis in that context is well established \citep[cf., for example,][\S\S2.9, 2.10]{batchelorFluidDyn}.  First, we note that

\begin{equation}
\int_{S_{b}} \en \cdot \nabla p^{(v)} \,\dee S = 0
\end{equation}
(via (\ref{eqpvbc2}) and Gauss's theorem).  Thus the analogous potential flow is that around a volume-conserving body, in which case (to within an arbitrary constant), $\phi(\ecks) \to 0$ as $|\ecks| \to \infty$.  More specifically, $\phi \sim |\ecks|^{-2}$ in three dimensions, and $|\ecks|^{-1}$ in two.  Hence $p^{(v)}$ exhibits the same dependencies.

\section{Viscous traction in terms of vorticity} \label{appvisctrac}
The viscous-stress tensor is given by \citep[\S3.3]{batchelorFluidDyn}

\begin{equation}
\tau_{ij} = \mu \left(\frac{\partial u_{i}}{\partial x_{j}}+\frac{\partial u_{j}}{\partial x_{i}}\right).
\end{equation}
The $i$th component of the force acting on the element $\dtwox$ of the (three-dimensional) body surface is $-\tau_{ij} n_{j} \dtwox$, with the negative sign arising because $\en$ is taken as pointing into the body.  Hence the result assumed in the main text is

\begin{equation}
-\mu\int_{S_{b}}\left(\frac{\partial u_{i}}{\partial x_{j}}+\frac{\partial u_{j}}{\partial x_{i}}\right)n_{j}\dtwox= \mu\int_{S_{b}}\left(\en\times\ohm\right)_{i}\dtwox.
\label{eqvtitom}
\end{equation}
Now the right-hand side can be written in terms of velocity gradients via index-based manipulation, yielding

\begin{equation}
\mu\int_{S_{b}}\left(\en\times\ohm\right)_{i}\dtwox = -\mu\int_{S_{b}}\left(\frac{\partial u_{i}}{\partial x_{j}} - \frac{\partial u_{j}}{\partial x_{i}}\right)n_{j}\dtwox.
\end{equation}
Thus the result in (\ref{eqvtitom}) can only be true if

\begin{equation}
\int_{S_{b}} \frac{\partial u_{j}}{\partial x_{i}} n_{j} \dtwox = 0.
\label{eqdujdxi}
\end{equation}
This can be shown via a variant of Stokes' theorem, given by \citet[\S1.4.4]{zangwillElectroDyn}:

\begin{equation}
\int\frac{\partial u_{j}}{\partial x_{i}}n_{j}\dtwox=\int\frac{\partial u_{j}}{\partial x_{j}}n_{i}\dtwox
\label{eqzang}
\end{equation}
over a closed surface.  Here, the right-hand side is zero by virtue of incompressibility.

Demonstrating (\ref{eqdujdxi}) in two dimensions is slightly more awkward, because the body surface is not closed.  Instead, consider the general form of the Stokes' theorem variant, which includes a line integral:

\begin{equation}
\left( \oint \mathbf{u} \times \dx \right)_i = \int\frac{\partial u_{j}}{\partial x_{j}}n_{i}\dtwox - \int\frac{\partial u_{j}}{\partial x_{i}}n_{j}\dtwox .
\end{equation}
Applying this identity to the three-dimensional surface created by extending the two-dimensional body contour normal to its plane, we find that the line integral consists of mutually cancelling elements, so (\ref{eqzang}) still applies.  Then, given the two-dimensionality, the area elements $\dtwox$ can be replaced by their line counterparts, $\dx$, so (\ref{eqdujdxi}) holds also in two dimensions.

\section{The missing viscous-stress component} \label{appmissvs}
The result claimed in the main text is

\begin{equation}
- \frac{1}{2}\mu\int_{S_{b}}\ecks\times\left(n_{j}\nabla\omega_{j}\right)\dtwox = \frac{1}{2}\tee ,
\end{equation}
where $\tee$ is given by (\ref{eqtee}).  The left-hand side can be manipulated to yield, in component representation,

\begin{equation}
\left[ -\frac{1}{2}\mu\int_{S_{b}}\ecks\times\left(n_{j}\nabla\omega_{j}\right)\dtwox \right]_k = -\frac{1}{2}\mu \epsilon_{kli} \int_{S_{b}} \frac{\partial}{\partial x_i} \left( x_l \omega_{j} \right) n_j \dtwox.
\end{equation}
Now the identity (\ref{eqzang}), with $u_{j}$ replaced by $x_l \omega_{j}$, can be applied to the right-hand side.  The resulting expression simplifies because vorticity is solenoidal ($\partial\omega_{j}/\partial x_j = 0$), so the upshot is

\begin{equation}
\left[ -\frac{1}{2}\mu\int_{S_{b}}\ecks\times\left(n_{j}\nabla\omega_{j}\right)\dtwox \right]_k = \frac{1}{2}\mu\int_{S_{b}} \epsilon_{kij} n_i \omega_j \dtwox .
\end{equation}
The right-hand side is recognisable as the component-representation form of $T_{k}/2$, confirming the claim.

\section{The relation between vorticity flux and slip growth} \label{appvfandslip}
In \S\ref{seccompvm} it was shown that the interpretation of the viscous vorticity flux as a boundary source responsible for the growth rate of the viscous vortex sheet at the body surface is erroneous; when invoked, it leads to a discrepancy between the traction-based and vorticity-moment representations of the force on the body.  Here we consider why this viewpoint is incorrect.

\begin{figure}
\begin{center}
\includegraphics{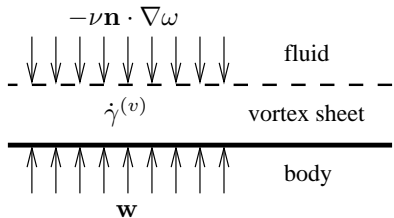}
\caption{{\label {vortflux}} The vorticity fluxes feeding the vortex sheet in the viscous component of the flow decomposition.}
\end{center}
\end{figure}

The key observation is that the viscous vorticity flux is a quantity defined in the fluid, and thus does not directly represent the body-surface vorticity source once the possibility of vortex-sheet generation is allowed.  Instead, the flux and the surface source combine to account for the vortex-sheet growth rate, as illustrated schematically in Fig.~\ref{vortflux}.  The body-surface flux is denoted by $\mathbf{w}$, and contributes to the vortex sheet from below.  The flux from the fluid is $-\nu\en\cdot\nabla\ohm$, and contributes from above.  Thus

\begin{equation}
\dot{\gam}^{(v)} = -\nu\en\cdot\nabla\ohm + \mathbf{w} .
\end{equation}
The assumption made by \citet{eldredgeReconciliation} corresponds to setting $\mathbf{w}$ equal to zero.  However, there is no theoretical justification for doing so; indeed the analysis of \S\ref{seccompvm} shows that it is inconsistent with the Navier-Stokes equation.  Instead, from (\ref{eqomflux}) and (\ref{eqomflux3d}), $\mathbf{w}$ is given by

\begin{equation}
\en \times \frac{\nabla p^{(v)}}{\rho}
\end{equation}
in two dimensions, and

\begin{equation}
\en \times \frac{\nabla p^{(v)}}{\rho} + \nu n_{j}\nabla\omega_{j} .
\end{equation}
in three.

This is not to say that Lighthill's original discussion is entirely wrong.  The first part, in which he considers the overall, non-decomposed, flow, and concludes that the surface acts as a vorticity source of strength $\nu\en\cdot\nabla\ohm$, is correct.  (In the overall flow there is no vortex-sheet growth, so the fluid-edge and surface vorticity fluxes are the same.)  The problem arises in the allocation of the surface flux between the inviscid and viscous aspects of the flow development.  Although Lighthill's description is qualitative, and hence not explicit, Eldredge's representation of it cannot be criticised as inaccurate.  On this interpretation, all the surface flux is responsible for the growth of the inviscid vortex sheet (because zero surface-flux contribution to the viscous sheet is implied, and there is no vorticity transfer from the fluid to the inviscid vortex sheet).  What has been shown here is that some of the surface flux must actually be apportioned to the viscous evolution.  Furthermore, the amount is not arbitrary; it is precisely specified by the equations governing the flow.

\bibliographystyle{jfm}
\bibliography{decFrcBodR1}

\end{document}